\documentclass[preprintnumbers,amsmath,amssymb,eqsecnum]{revtex4}
\usepackage{epsfig}

\newcommand{\bea}{\begin{eqnarray}}
\newcommand{\ena}{\end{eqnarray}}

\newcommand{\be}{\begin{equation}}
\newcommand{\ee}{\end{equation}}
\newcommand{\ba}{\begin{array}}
\newcommand{\ea}{\end{array}}

\begin{document}

\title{\bf Age constraints on the Agegraphic Dark Energy Model }
\vskip.5in

\author{Yi Zhang$^{1,2,}$\footnote{Email: zhangyi@itp.ac.cn}, Hui
Li$^{1,2,}$\footnote{Email: lihui@itp.ac.cn}, Xing
Wu$^{3,}$\footnote{Email: wxxwwxxw@mail.bnu.edu.cn}, Hao
Wei$^{4,}$\footnote{Email: haowei@mail.tsinghua.edu.cn}, Rong-Gen
Cai$^{1,}$\footnote{Email: cairg@itp.ac.cn} }
\address{
 $^{1}$Institute of Theoretical Physics, Chinese Academy of
Sciences P.O. Box 2735, Beijing 100080,
China \\
 $^{2}$Graduate
University of the Chinese Academy of Sciences,
Beijing 100049, China \\
 $^{3}$Department of Astronomy, Beijing Normal University,
Beijing 100875, China \\
 $^{4}$Department of Physics and Tsinghua Center for
 Astrophysics, Tsinghua University, Beijing 100084, China }


\begin{abstract}
We investigate the age constraint on the agegraphic dark energy
model by using two old galaxies (LBDS $53W091$ and LBDS $53W069$)
and the old high redshift quasar APM $08279+5255$. We find that the
agegraphic dark energy model can easily accommodate LBDS $53W091$
and LBDS $53W069$. To accommodate APM $08279+5255$, one can take the
reduced Hubble parameter as large as $h=0.64$, when the fraction
matter energy density $\Omega_{m0}\approx 0.22$.
\end{abstract}
 \maketitle

\section{Introduction}\label{sec1}

The observational data strongly suggest that the universe is
accelerating today\cite{r1,r2,r3,r4,r5,r6,r7,r8}; as a consequence,
the study of dark energy has been one of the most active topics in
modern cosmology. The simplest candidate for dark energy is the
famous cosmological constant which, however, is plagued with the
so-called ``cosmological constant problem'' and ``coincidence
problem''~\cite{r9}. Some dark energy models have been proposed with
the dynamical scalar field(s), and some others by means of plausible
quantum gravity arguments; the former refers to the well-known
quintessence~\cite{r10,r11}, phantom~\cite{r12,r13,r14},
k-essence~\cite{r15} and quintom models~\cite{r16,r17,r18,r19},
hessence~\cite{r20}, etc, while the latter contains holography dark
energy~\cite{r21}, for instance.

As an important next step, these theoretical models have to be
confronted by observational data. Indeed, a lot of observational
constraints on these models have been carried out by using
observational data, for example, from SNe, CMB, large scale
structure (LSS), etc.  Recently a kind of new constraints has
attracted a lot of attention: the age of some old high redshift
objects (OHROs) as a constraint on the cosmological model. The basic
idea is that these OHROs can not be older than the universe itself.
In the literatures, one usually uses the age of three OHROs to
constrain some theoretical models: two of them are old galaxies
(LBDS $53W091$ and LBDS $53W069$) and the other is a high redshift
quasar (APM $08279+5255$). The relevant data are listed in Table
\ref{tab1}.
\begin{table}[ht]
\caption{The redshift and age of three old objects.}
\begin{center}
  \begin{tabular}
  {r@{\hspace{5mm}}r@{\hspace{5mm}}cc}
  \hline \hline
     & Name & Redshift &\,\,\,Age \\
    \hline
     &LBDS 53W091 & $\,\,\,z=1.55$& 3.5~Gyr   \\
     &LBDS 53W069 &  $\,\,\,z=1.43$ & 4.0~Gyr  \\
     & APM 08279+5255 & $\,\,z=3.91$& 2.0-3.0~Gyr or 2.1~Gyr\\
    \hline
  \end{tabular}
\end{center}\label{tab1}
\end{table}
where $z$ is the redshift with the definition $z=a^{-1}-1$ (assuming
today's scale factor $a_0=1$ ). Those three OHROs are used
extensively in the group of the old objects in our universe as
today's observation mentioned. LBDS $53W091$ \cite{r24,r25} was
$3.5~Gyr$ old at $z=1.55$. LBDS $53W069$ \cite{r26} was 4.0~Gyr at
$z=1.43$. The age of APM $08279+5255$ is in debate: one
method~\cite{r27,r28} shows its age between 2.0-3.0~Gyr at $z=3.91$
, while the other method~\cite{r29} shows that it was 2.1~Gyr old at
the same redshift. We use $T=2.0Gyr$  as the age of the APM
08279+5255($z=3.91$). Because the $T=2.0Gyr$ is the lowest limit of
the age of APM 08279+5255 ($z=3.91$), it is the loosest constraint
of APM 08279+5255($z=3.91$) on the age of the universe.

It turns out that the constraints on the age of the universe are not
easy to satisfy. Taking the matter-dominated flat FRW universe, for
example, its age $T$ reads:
 \be \label{Eq3} T = \frac{2}{3}H_0^{-1}(1 + z)^{-3/2},\ee
where $H$ is the Hubble parameter and the subscript ``$0$" denotes
today's value of the corresponding quantity at redshift $z=0$. We
 define a dimensionless parameter $h$ for convenience with
$H_{0}=100h \, km\cdot s^{-1}\cdot Mpc^{-1}$. According to
$(\ref{Eq3})$, the flat matter-dominated FRW model can be ruled out
unless $h<0.48$. Not only does the flat matter dominated FRW model
have the problem of being compatible with the observational
age-redshift relation of OHROs, but also the closed FRW matter
dominated model.  The age problem becomes even more serious when we
consider the age of the universe at high redshift. The age problem
is one of the reasons that we need an accelerated expansion of the
universe today. When some dark energy component is introduced to the
universe, it is shown that most dark energy models can only
accommodate data of LBDS $53W091$ ($z=1.55$) and LBDS $53W069$
($z=1.43$), but unfortunately cannot be compatible with APM
$08279+5255$($z=3.91$). The list is long: the dark energy models
with different EoS parameterizations~\cite{r32,r33}, the generalized
Chaplygin gas~\cite{r34}, the $\Lambda(t)$CDM model~\cite{r35}, the
model-independent EoS of dark energy~\cite{r36}, the scalar-tensor
quintessence~\cite{r37}; the $f(R)$ model~\cite{r38}, the DGP
braneworld model~\cite{r39,r40}, the power-law parameterized
quintessence model~\cite{r41}, holographic dark energy ~\cite{r51}
and so on. In particular, the most famous and WMAP most favored
model $\Lambda$CDM is also included in the list~\cite{r29,r30,r31}.
This gives rise to the so-called  age crisis in dark energy
cosmology.

  More recently, a new dark energy model, named agegraphic dark energy, has been
  proposed~\cite{r23}, which takes into account the
uncertainty relation of quantum mechanics together with the
gravitational effect in general relativity. One has the so called
K\'{a}rolyh\'{a}zy relation $\delta t=\beta t_{p}^{2/3}t^{1/3}$
\cite{r22}, and energy density of spacetime
fluctuations~\cite{in1,in2,in3}
 \be \label{Eq1}
 \rho_{q}\sim
\frac{1}{t_{p}^{2}t^{2}}, \ee
 $\beta$ is a numerical
factor of order one, and $t_{p}$ is the Planck time. The agegraphic
dark energy model assumes that the observed dark energy comes from
the spacetime and matter field fluctuations in the universe. The
dark energy has the form (\ref{Eq1}) and $t$ is identified with the
age $T$ of the universe~\cite{r23}
 \be \label{Eq2}
\rho_{q}=\frac{3n^{2}M_{pl}^{2}}{T^{2}},
 \ee
  where $M_{pl}$ is the reduced Planck mass and the constant
   $n^{2}$ has been introduced for representing some unknown theoretical
uncertainties. Both in the radiation-dominated and matter-dominated
epoches, the energy density of the agegraphic dark energy just
scales as $\rho_q\sim t^{-2}$ tracking the dominant energy
component. Moreover, the model can also make the late-time
acceleration. For the further development of the model, see
\cite{Wei1,Wu,Wei2}.

In this paper we will use those three ORHOs to constrain the
agegraphic dark energy model and see whether resulting constraints
are compatible with constraints coming from other observational
data.  In the next section we introduce the method used to constrain
the model. In Sect.~3 we will give the result on the age problem in
the agegraphic dark energy.  Sec.~4 will be devoted to the
conclusion.


\section{The Method}
\label{sec2}

\subsection{The age of the universe versus redshift}
We consider a spatially flat FRW universe which contains the
agegraphic dark energy and the pressureless (dark and dust) matter.
The Friedmann equation is given by
 \be \label{Eq4}
 3M_{pl}^2H^2=\rho_m+\rho_q,
\ee where $\rho_m$ and $\rho_q$ are the energy density of the
 pressureless matter and the agegraphic dark energy
 respectively. We
 assume there is no direct coupling between them and each of them satisfies the continuity equation separately,
 \be \label{Eq5} \dot{\rho}_{m}=-3H\rho_{m},\ee
\be \label{Eq6}\dot{\rho_{q}}=-3H(\rho_{q}+p_{q}).\ee Taking
derivative of Eq. (\ref{Eq2}) with respect to $t$, we obtain
 \be \label{Eq7} \dot{\rho_{q}}=-2H\rho_{q}\sqrt{\Omega_{q}},\ee
where $\Omega_{q}=\rho_{q}/3M_{pl}^2H^2$ is the fractional energy
density of the dark energy component. One can also define
$\Omega_{m}=\rho_{m}/(3M_{pl}^2H^2)$ as the fractional energy
density of the pressureless matter. The Friedmann equation then can
be rewritten as
 \be \label{Eq8}
 \Omega_{q}+\Omega_{m}=1.
 \ee
Furthermore, taking derivative of the above formula with respect to
$z$ and considering the relationship $dz=-H(1+z)dt$, we reach
 \be \label{Eq9}
 \frac{d\Omega_{q}}{dz}=-(1+z)^{-1}\Omega_q
 (1-\Omega_q)\left(3-\frac{2}{n}\sqrt{\Omega_{q}}\right).
 \ee
This is the evolution equation which encodes the main information of
the FRW cosmology. Given an initial $\Omega_{q}$ at certain $z$, we
may evaluate $\Omega_{q}$ at any specific redshift in terms of the
model parameter $n$.


\begin{figure}[ht]
\begin{center}
\includegraphics[width=9cm]{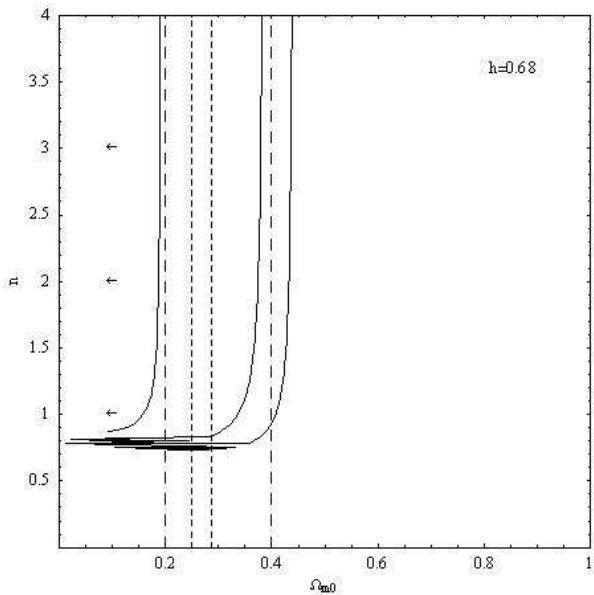}
\caption{\label{fig1} The three solid lines are,
 from left to right, contours
 $T_z(3.91)=T_{zobj}(3.91)$, $T_z(1.43)=T_{zobj}(1.43)$ and
 $T_z(1.55)=T_{zobj}(1.55)$. Only using the age constraint, it is obvious that the allowed parameter pairs
 $(\Omega_{m0}, n)$ should lie in the left common region of these three contours, as is indicated by
 the arrows. For a cross-check procedure with other observations, the WMAP3 bound $\Omega_{m0}=0.268\pm 0.018$~\cite{r5} is indicated by two
 short-dashed lines and the model-independent cluster estimate
 $\Omega_{m0}=0.3\pm 0.1$~\cite{r50} is indicated by two
 long-dashed lines. Here, we have used the reduced Hubble constant $h=0.68$.} \label{fig:68}
\end{center}
\end{figure}



\begin{figure}[ht]
\begin{center}
\includegraphics[width=9cm]{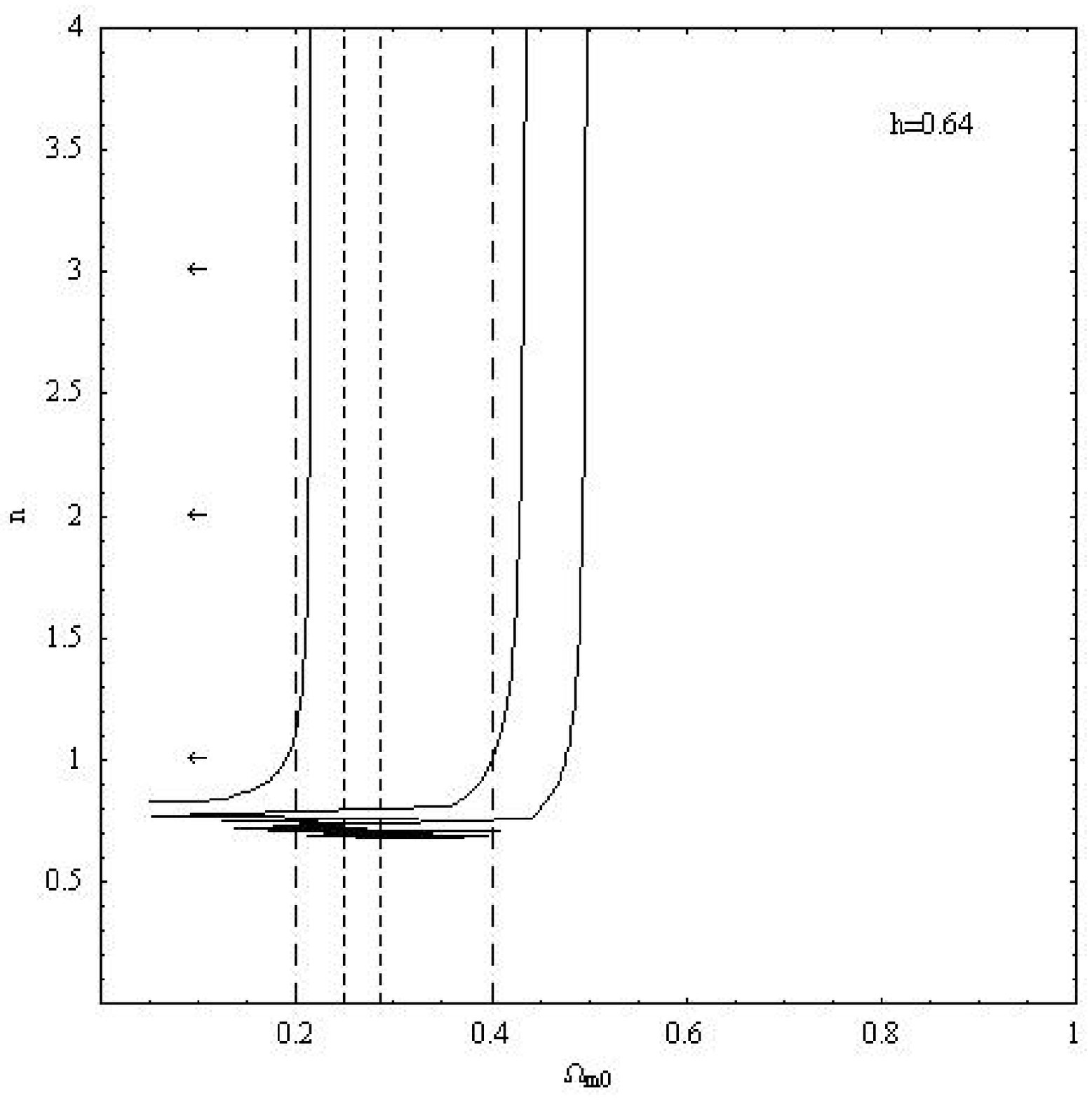}
\caption{\label{fig2} The same as in Fig.~\ref{fig1},
 except for $h=0.64$.}
\end{center}
\end{figure}


The age of our universe at redshift $z$ is given by
 \be \label{Eq10}
 T(z)=\int^{a}_{0}\frac{da}{aH}=\int_z^\infty\frac{d\tilde{z}}{(1+\tilde{z})H(\tilde{z})}.
 \ee\
$T(z)$ means the age of the universe at a certain redshift $z$. It
is convenient to introduce a dimensionless age parameter~\cite{r33}
 \be
\label{Eq11} T_z(z)= H_0 T(z)=\int_z^\infty
 \frac{d\tilde{z}}{(1+\tilde{z})E(\tilde{z})}\,,
\ee
 where $E(z)\equiv H(z)/H_0$. Considering the evolution of
matter, $\rho_{m}=\rho_{m0}(1+z)^{-3}$, we can easily get:
 \be \label{Eq12}
 E(z)=\left[\frac{\Omega_{m0}(1+z)^3}{1-
 \Omega_q}\right]^{1/2},
\ee
 So if $\Omega_{m0}$ is fixed, the value of $\Omega_{q0}$ is
 fixed by Eq.(\ref{Eq8}). We can get the values of $\Omega_{q}$ by Eq.(\ref{Eq9}),
 $E(z)$ by Eq.(\ref{Eq12}), $T_{z}(z)$ by Eq.(\ref{Eq11}) at any redshift with only a
 model parameter $n$.

 At any redshift, the age of our universe should not be smaller than
the age of the OHROs, namely
 \be \label{Eq13}
 T_z(z)\geq T_{zobj}= H_0 T_{obj} \ee
where $T_{obj}$ is the age of the corresponding OHROs. If we fixed
$T_z(z)=T_{zobj}$ , from the analysis before, we can get the value
of $n$ with the initial $\Omega_{m0}$ fixed. Here we can not
forget the parameter $H_0$ in the equation $T_{zobj}= H_0 T_{obj}$
which should also be fixed in advance. We discuss the choice of
the parameters in the next subsection.

Let's make a short summary. If we had the value of $H_{0}$, we
know the value of $T_{zobj}$ by Eq.(\ref{Eq13}). And given an
initial $\Omega_{m0}$, it is equivalent to give an initial
$\Omega_{q0}$, then we can know $\Omega_{q}$ by (\ref{Eq9}),
$E(z)$ by Eq.(\ref{Eq12}), and $T_{z}(z)$ by Eq.(\ref{Eq11}) with
the model parameter $n$. Using the relation $T_z(z)=T_{zobj}$, we
can get a model parameter $n$.  If we let $\Omega_{m0}$ run in a
range, there is a curve showing the relation between the different
$\Omega_{m0}$ and its corresponding minimal estimation of $n$.


\begin{figure}[ht]
\begin{center}
\includegraphics[width=9cm]{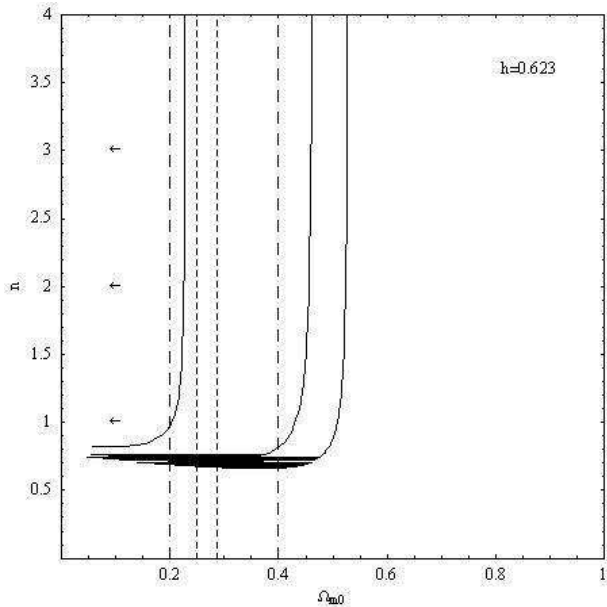}
\caption{\label{fig3} The same as in Fig.~\ref{fig1},
 except for $h=0.623$.}
\end{center}
\end{figure}


\subsection{The choice of the parameter}\label{Subsec2}
 There are various observational methods such as SN~Ia, CMB, LSS,
which could give out many data on the cosmological parameters. We
will use the scope of some cosmological parameters  provided by them
to
 test the age problem in the agegraphic dark
energy model with respect to three OHROs listed in Table \ref{tab1}.

As stated in the previous subsection, one should first choose an
$\Omega_{q0}$ or $\Omega_{m0}$ as an initial condition. The loosest
$\Omega_{m0}$ value is $\Omega_{m0}=0.3\pm 0.1$  from
model-independent cluster estimation~\cite{r50}. A tighter WMAP3
bound is
 $\Omega_{m0}=0.268\pm 0.018$~\cite{r5} with the combined
 constraint from the latest SNe~Ia, galaxy clustering and CMB
 anisotropy.

 The choice of Hubble constant is not a direct one. In the literatures,
  the reduced Hubble constant $h=0.72\pm0.08$
 of Freedman {\it et al.}~\cite{r43} has been used extensively.
This Hubble constant seems too high to explain away the age problem.
And many authors also argue for
 a lower Hubble constant, for instance, $h=0.68\pm 0.07$
 at $2\,\sigma$ confidence level in~\cite{r49}.
In the past few years, it has been also argued that there exits
 systematic bias in the result of Freedman {\it et al.}~\cite{r43}.
 Sandage and collaborators advocate a lower Hubble constant in
 a series of works~\cite{r44,r45,r46,r47,r48}, and their final result
 reads $h=0.623\pm 0.063$~\cite{r48}.
 We will take the parameter values of $h=0.68$ (the center value in~\cite{r49} ),
 $h=0.64$ (the lower limit of $h=0.72\pm
0.08$), $h=0.623$ (the center value after eliminating the alleged
systematic bias), $h=0.59$ (a useful mediate value), $h=0.56$ (the lower limit of $h=0.623\pm 0.063$ )
in our analysis in turn.\\

\begin{figure}[ht]
\begin{center}
\includegraphics[width=9cm]{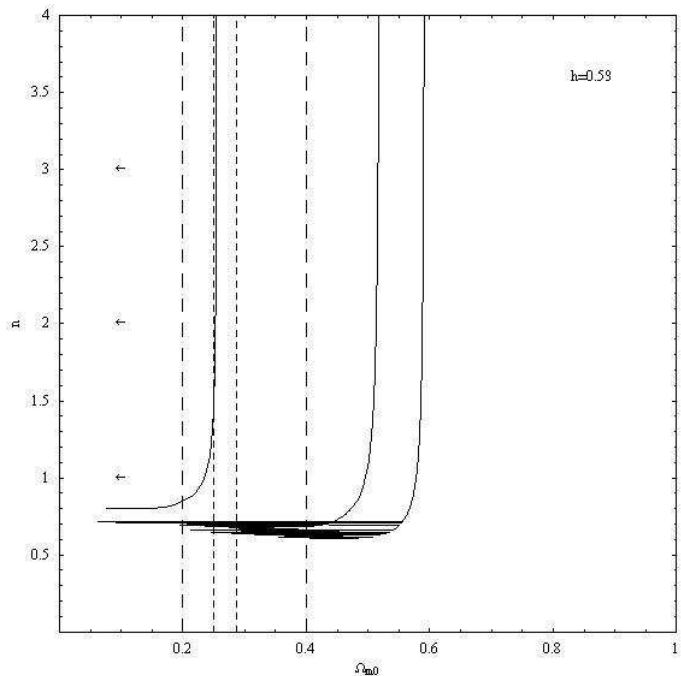}
\caption{\label{fig4} The same as in Fig.~\ref{fig1},
 except for $h=0.59$.}
\end{center}
\end{figure}


\begin{figure}[ht]
\begin{center}
\includegraphics[width=9cm]{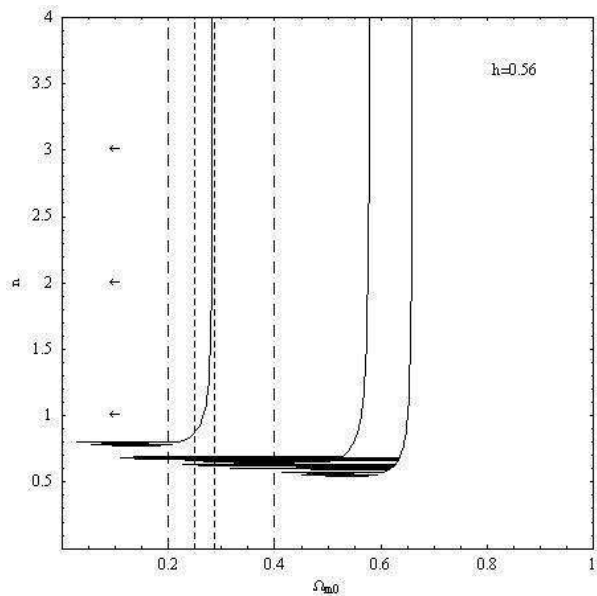}
\caption{\label{fig5} The same as in Fig.~\ref{fig1},
 except for $h=0.56$.}
\end{center}
\end{figure}


\section{Results}\label{sec3}

We show the numerical results in Fig.~\ref{fig1}-\ref{fig5} for
different Hubble parameters. From Eq.~(\ref{Eq12}), the lower
$\Omega_{m0}$, the lower $E(z)$ is, and then from Eq.~(\ref{Eq11}),
the larger $T_{z}(z)$ is, so the allowed parameter space must be
constrained to the left by the three curves $T_{z}(z)> T_{zobj}$.
The leftmost curve $T_{z}(3.91)> T_{obj}(3.91)$ gives the most
stringent bound and gives consequently the allowed parameter space,
which corresponds to the common left regions of these three
contours, as the arrows indicate.  We  see from the figures that
when $h$ decreases from $0.68$ to $0.56$, the allowed parameter
space expands horizontally to the right and begins to cover the
regions which are required by cluster estimation of
$\Omega_{m0}=0.3\pm 0.1$ and WMAP3 $\Omega_{m0}=0.268\pm 0.018$,
respectively. The constraints given by LBDS $53W091$ ($z=1.55$) and
LBDS $53W069$ ($z=1.43$) can be easily satisfied in the agegraphic
dark energy model, as many other dark energy models. The most
difficult one to accommodate is APM 08279+5255. In Table
\ref{table2} we list the exact result for APM 08279+5255 curves in
Fig. \ref{fig1}-\ref{fig5}:

\begin{table}[ht]
\caption{The range of fraction matter energy density, which can
accommodate the old quasar APM 08279+5255, with different Hubble
parameter.}
\begin{center}
  \begin{tabular}{cccc}
  \hline \hline
     $h$& $\Omega_{m0}(0.2,0.4)$ & $\Omega_{m0}(0.22,0.4)$& $\Omega_{m0}(0.25,0.286)$\\
    \hline
     $0.56$ & Ok &   Ok  & Ok \\
     $0.59$ & Ok &   Ok & Ok \\
     $0.623$ &  Ok &  Ok & No  \\
     $0.64$ & Ok & Marginal & No\\
    $0.68$ & No & No   &No  \\
    \hline
\end{tabular}
\end{center}
\label{table2}
\end{table}

If $\Omega_{m0}=0.3\pm 0.1$,  Fig.\ref{fig2} shows that one can take
the Hubble parameter as large as $h=0.64$ in our model. When
$h=0.59$, the parameter scope coming from WMAP3 begins to be
reached, while $h=0.56$ makes the allowed region cover nearly the
whole interval $\Omega_{m0}=0.268\pm 0.018$. We therefore conclude
that our model alleviate the age crisis.

We can make a simple comparison among some dark energy models. The
WMAP most favored $\Lambda CDM$ model can not accommodate APM
08279+5255 ($z=3.91$) unless the Hubble parameter is taken as low as
$h=0.58$~\cite{r29}, and $\Lambda(t)$CDM model considered in
\cite{r35} can not change this conclusion. In Table~\ref{tab3}, We
list some results for $\Lambda(t)CDM$ model, $\Lambda$CDM/DGP model,
holographic dark energy model and agegraphic dark energy model, in
order to accommodate the APM 08279+5255.
\begin{table}[ht]
\caption{A rough comparison of some dark energy models}
\begin{center}
\begin{tabular}{rccc}
\hline \hline
Name& $\Omega_{m0}$ & $h$& $z$\\
\hline
$\Lambda(t)$CDM\cite{r35} & 0.2 &   0.64  &$ >5.11$ \\
Holography DE \cite{r51}&  $\sim$ 0.2&  0.64 & 3.91  \\
$\Lambda$CDM/DGP \cite{r29} & 0.23 & 0.58 & 3.91\\
Agegraphic DE & $\sim$ 0.22 & 0.64& 3.91\\
\hline
\end{tabular}
\label{tab3}
\end{center}
\end{table}

According to \cite{r35}, even when the present energy density
 of matter takes a value as low as $\Omega_{m0}=0.2$ and
$h=0.64$, one still finds $z>5.11$, which is clearly incompatible
with the observation $z=3.91$. When $h=0.64$, $z=3.91$, one could
have $\Omega_{m0}\sim0.2$~\cite{r51}, in the holography dark energy,
while in the agegraphic dark energy model, the same parameter
evaluation tells us $\Omega_{m0}\sim0.22$.

In addition, let us mention here three features of the behavior on
the solid lines $T_{z}(z)=T_{zobj}$ in the figures. (1) at the
bottom of the curves with low $\Omega_{m0}$ and small $n$, the
curves have an oscillation behavior.  (2) at the middle of the
curves, as $\Omega_{m0}$ increases,  $n$ increases, and the curves
in $(n-\Omega_{m0})$ plane, go from the left-bottom to the
right-top. (3) at the top of the curves, $n$ increases fast but
$\Omega_{m0}$ nearly keeps unchanged.

\begin{figure}[ht]
\begin{center}
\includegraphics[width=9cm]{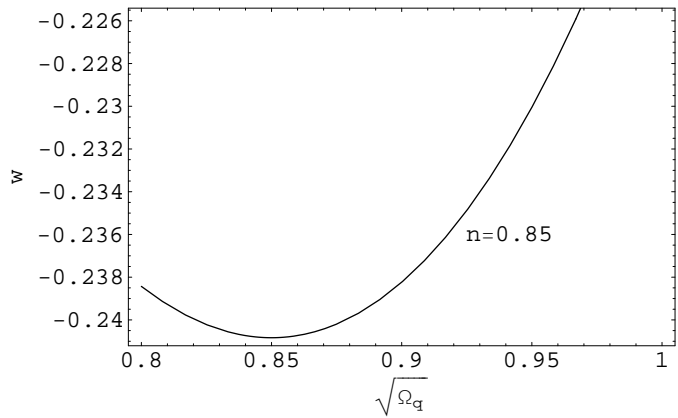}
\caption{\label{fig6}The relationship between $w$ and
$\sqrt{\Omega_{q}}$ when $n=0.85$, one value of $w$ may correspond
to two different values of $\Omega_{q}$, and then one value of $n$
may result in two different values of $\Omega_{q}$.}
\end{center}
\end{figure}

We may give a possible explanation for the oscillation behavior from
the state of parameter (EoS) $w$ of the universe. It reads
 \be
 w=\frac{p_{m}+p_{q}}{\rho_{m}+\rho_{q}}=\Omega_{q}\left(-1+
 \frac{2\sqrt{\Omega_{q}}}{3n} \right ).\label{Eq14} \ee
Taking derivative of Eq.(\ref{Eq14}) with respect to
$\sqrt{\Omega_{q}}$, one has \be \frac{d
w}{d\sqrt{\Omega_{q}}}=-2\sqrt{\Omega_{q}}(1-\sqrt{\Omega_{q}}/n).\ee
We can find that, as $\Omega_{q}$ increases from zero, $w$
monotonously decreases first and then increases, therefore the
transition could happen at $\sqrt{\Omega_{q}}=n$ if $n<1$. That is,
if $n$ was too low while $\Omega_{m0}$ was low too, one value of $n$
may correspond to two $\Omega_{q0}$ or $\Omega_{m0}$ equivalently;
as a consequence, the oscillation of the curves at the left-bottom
of $(n-\Omega_{m0})$ parameter space appears. This argument may be
illustrated in Fig. \ref{fig6}. In fact, the region with the
oscillation behavior should be regarded as unphysical one since in
order to have an accelerated expansion for the universe, the model
requires $n>1$~\cite{r23}.

We see from the figures that $n$ increases very quickly when
$\Omega_{m0}$ gets large. Finally $n$ becomes insensitive to
$\Omega_{m0}$. Thus we can hardly get the upper limit of the
parameter $n$ since the curve is too straight to leave from the
allowed parameter region as the parameter $n$ increases. However
the contour given by $T_{z}(z)=T_{zobj}$ is enough to give
 the age constraint we need. We can know from figures whether the resulting range of
 $\Omega_{m0}$ is in the range given by  other
 observational constraints. In the figures
 the WMAP3 bound $\Omega_{m0}=0.268\pm 0.018$~\cite{r5} is indicated by two
 short-dashed lines and the model-independent cluster estimation
 $\Omega_{m0}=0.3\pm 0.1$~\cite{r50} is indicated by two
 long-dashed lines.

\section{Conclusion}

The relationship between the red-shift $z$ and the cosmic age $t$
 contains a lot of information about the evolution of the
universe. In this paper we have investigated the age constraint on
the agegraphic dark energy model by using the observational data
for two old  galaxies (LBDS $53W091$ with $z=1.55$  and LBDS
$53W069$ with $z=1.43$) and an old high redshift quasar (APM
$08279+5255$ with $z=3.91$). As most dark energy models, the
agegraphic dark energy model can easily accommodate these two old
galaxies. In order to accommodate the old quasar, one has to take
a little lower Hubble parameter $h=0.64$ when $\Omega_{m0}\approx
0.22$. Although the behavior looks slightly better than some other
dark energy models, the age crisis is still there, unless the
current Hubble parameter has indeed a lower value than the best
fitting value of WMAP3, for example, $h=0.59$ as advocated by
Sandage and collaborators.

Finally we stress here that the numerical results are obtained
through integrating the equation (\ref{Eq9}) by imposing an initial
condition, for example, $\Omega_q=0.73$ at redshift $z=0$.  As
stressed in \cite{r23}, the equation (\ref{Eq9}) not only holds for
the form $T=\frac{n}{H\sqrt{\Omega _q}}$, but also for another form,
$T'=T+\delta=\frac{n}{H\sqrt{\Omega_q}}$. And the constant $\delta$
can be obtained by $\delta =\frac{n}{H\sqrt{\Omega_q}}-\int^a_{0}
\frac{da}{Ha}$. The integration from $z=0$ to $z=\infty$ does not
guarantee the constant $\delta$ vanishes. As a result, the energy
density could have a form $\rho_q=
\frac{3n^2M^2_{pl}}{(T+\delta)^2}$ in a general case. In addition,
it would be of great interest to study the age constraint on the new
model of agegraphic dark energy~\cite{Wei2}, where a conformal time
scale is introduced to the energy density (\ref{Eq2}), instead of
the cosmic age.


\section*{ACKNOWLEDGMENTS}
We are grateful to Bin Hu for useful discussions. We also thank
Fu-qiang Xu and  Ding Ma for kind help. The work was supported in
part by a grant from Chinese Academy of Sciences (No. KJCX3-SYW-N2),
and by NSFC under grants No.~10325525, No.~10525060 and
No.~90403029.


\end{document}